\documentstyle[psfig,twocolumn,aps]{revtex}
\begin{document}
\draft
\wideabs{\title{
Upper critical field peculiarities of superconducting 
YNi$_2$B$_2$C and LuNi$_2$B$_2$C}
\author{ 
S.V.\ Shulga \cite{shu}, S.-L.\ Drechsler\cite{dre}, G.\ Fuchs, and K.-H.\ 
M\"uller}
 \address{ Institut f\"ur Festk\"orper- und Werkstofforschung Dresden e.V.,
 Postfach 270016, D-01171 Dresden, Germany }

\author{K.\ Winzer, M.\ Heinecke, and K.\ Krug } 
\address{ 1. Physikalisches Institut, Universit\"at G\"ottingen, 
Bunsenstrasse 9,
D-37073 G\"ottingen, Germany}
\date{7 October 1997}
\maketitle
\begin{abstract}
We present  new upper critical field $H_{c2}(T)$ data in a  broad 
temperature region 0.3K $\leq T\leq T_c$
for LuNi$_2$B$_2$C and 
YNi$_2$B$_2$C single crystals
with  well characterized low 
impurity scattering rates.
 The absolute values for all $T$, in particular  $H_{c2}(0)$,
  and  the sizeable 
positive curvature (PC) of $H_{c2}(T)$ at high and intermediate $T$ 
are   explained quantitatively within an effective
 two-band model.
The failure of the  isotropic single band approach is discussed
 in detail. Supported by de Haas van Alphen data, 
  the  superconductivity reveals direct insight 
 into details of the electronic structure. The observed maximal  
 PC  near $T_c$ gives strong
 evidence for  clean limit type II superconductors.
  \end{abstract}
}
\narrowtext
 The  discovery\cite{cava94,nagarajan94} of superconductivity in
transition metal borocarbides has generated large interest due to
 their
relatively high transition temperatures $T_c\sim $ 15 to 23 K and 
due to the relation between  the mechanisms of superconductivity
 in these compounds, in cuprates, and in ordinary transition
 metals. Another  highlight is the coexistence 
of magnetism and superconductivity in  some of these compounds containing 
 rare earth elements \cite{cho95,evertsmann96,fuchs95}.
A   
study of the non-magnetic compounds such as {\it L}Ni$_2$B$_2$C, with  
{\it L}=Lu,Y,Th,Sc \cite{lai95},
is a prerequisite for the  
understanding of  their magnetic counterparts.
Experimental data for LuNi$_2$B$_2$C \cite{metlushko97}
 demonstrate beside a maximal positive curvature (PC) of $H_{c2}(T)$ near
  $T_c$,
  observed also 
 for
 YNi$_2$B$_2$C \cite{michor95,evertsmann96,rathnayaka97},
   a weak  
$T$-dependent  anisotropy within the tetragonal basal plane 
and a $T$-independent out-of-plane anisotropy of the upper critical field 
$H_{c2}$.
Both anisotropies have been described\cite{metlushko97} in terms of  nonlocal 
corrections
to the Ginzburg-Landau (GL) equations. In this picture  the PC
  of $H^c_{c2}$ 
($\vec{H} \| $ to the tetragonal c-axis) is
 caused, almost purely, by the  basal plane  anisotropy. 
 However, it should be noted 
that the reported anisotropy of $H_{c2}$ for YNi$_2$B$_2$C is
significantly smaller than for LuNi$_2$B$_2$C
\cite{metlushko97,rathnayaka97,dewilde97}
 whereas its PC is comparable or even
 larger. Further 
explanations of the unusual PC of $H_{c2}(T)$, such as  
quasi-2D
fluctuations  \cite{bahcall95}, are excluded by the 
underestimation of $H_{c2}(T)$ at low-$T$ \cite{rathnayaka97} and the
observed  weak anisotropy.
 The quantum critical point scenario 
 \cite{kotliar96} as well as  the bipolaronic one 
 \cite{alexandrov93} can be disregarded because 
 the  slope of $H_{c2}(T)$ decreases for $T$$ \to$0 
(see Fig.\ 1). 
Local density approximation (LDA)
band structure calculations\cite{pickett94,lee94} predict a 
nearly isotropic electronic structure with rather complicated bands
 near the
Fermi level $E_F$. However, in analyzing the superconductivity 
in terms of an isotropic {\it single}-band (ISB) Eliashberg
model, the multi-band character and the  anisotropic Fermi
 surface  have been widely ignored so far.

Here we present and analyze theoretically new data
 of $H_{c2}(T)$ in a broad 
 interval  
  0.3K$\leq T\leq T_c$ 
  for  high purity 
LuNi$_2$B$_2$C and YNi$_2$B$_2$C single crystals.
We  show that typical  
features of both compounds,
such as  $H_{c2}(0)\sim$ 8 to 10 
T and the unusual PC of $H_{c2}(T)$ for
 $T\stackrel{>}{\sim}0.5T_c$,  
 cannot in any way be explained   
consistently with the normal state properties
within the ISB approach.
Instead we propose a two-band model (TBM) approach. To clarify its 
 relationship  to the extended saddle-point model
 \cite{abrikosov97} which also predicts a PC is beyond the scope
  of our letter.
 
 Platelet shaped LuNi$_2$B$_2$C and YNi$_2$B$_2$C
single crystals with a mass of $\sim$ 1 mg were grown 
by a high-temperature flux technique with Ni$_2$B as flux.
 The 
  values
of
$T_c$, 16.5 K and 15.7 K, have been determined by low-
\begin{figure}
\psfig{figure=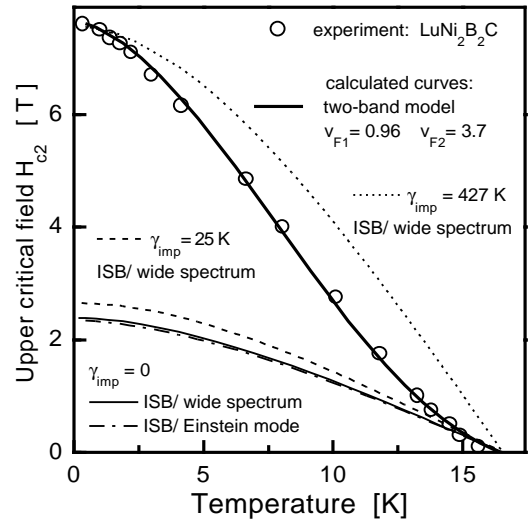,width=8.0cm,height=7.9cm}
\caption{ Experimental data for $H_{c2}(T)$ of
LuNi$_2$B$_2$C (magnetic field $\vec{H} \|$ the c-axis)
 compared with theoretical curves: (i) the isotropic
single band (ISB) model with Fermi velocity 
 $v_F$=2.76$\cdot$10$^{7}$ cm/s and various
  impurity scattering rates
$\gamma_{imp}$ explained in the text and the 
legends and (ii) the two-band model (TBM) with
 $v_{Fi}$ (i=1,2)
  in units of
 10$^{7}$ cm/s.}
 \label{fig1}
 \end{figure}
\noindent
field ac susceptibility $\chi (T)$
 with
 transition widths $\Delta T_c$ = 0.2 K.
The upper critical
 field $H^c_{c2}(T)$ along the  c-axis, shown in Figs.\ 1 and 2,  
 has been measured
 resistively for
  fixed $T$
  adopting the midpoint criterion:
$\rho (H_{c2},T)$=0.5$\rho$($H$=0,$T$=17K)$\equiv 0.5\rho_n$. The 
 transition width  $\Delta H$=$[H(\rho$=0.9$\rho_n)-H(\rho$=0.1$\rho_n )]$ 
 increases up to 0.75 T (1.5 T for LuNi$_2$B$_2$C) at $T<$ 0.5 K  
starting from a nearly constant value of 0.3 T
at $T>$4 K (8 K for LuNi$_2$B$_2$C).
 The  low residual resistivity 
$\rho(0) \approx \rho_n$=2.5$\mu \Omega$cm and the ratio
$\rho(300{\rm K})/\rho_n$ =43 (27 for LuNi$_2$B$_2$C), together with 
the
observations of magnetoquantum oscillations
 \cite{heinecke95,goll96,nguyen96} indicate 
 a high quality and a low impurity content of our samples.
This suggests that we are in  the clean limit  in terms of the
traditional theory of type II superconductors\cite{remark0}. 
In this limit one  has to consider the electronic structure in more  detail.  
We restrict ourselves to 
an effective two-band model \cite{remark2} 
which, especially on the simple
   BCS-level, has a long history \cite{moskalenko73}.
     Due to the neglect of strong coupling effects, a BCS-like theory
    is not expected to describe real superconductors quantitatively.
    Such effects must be studied within the Eliashberg theory
    \cite{prohammer87,entel76,allen76,langmann92,carbotte90}.  To 
  calculate $H_{c2}$ we have solved numerically the corresponding linearized 
equations of Ref.\ 23, 
\begin{eqnarray}
\tilde{\omega}_i(n) & = & \omega_n+\pi T \sum_{j,m}  [\lambda_{i,j}(m-n)
+ \nonumber \\
       &  & +  \delta_{mn}(\gamma_{imp;i,j}+ \gamma^s_{imp;i,j})/2\pi T] \mbox{sgn}(\omega_m) , \\
\tilde{\Delta}_i(n) & = &\pi T \sum_{j,m}  [\lambda_{i,j}(m-n) - \mu^{*}
\delta_{ij}\theta(\omega_c-\mid\omega_m\mid)+ \nonumber \\
       &  & +\delta_{mn}(\gamma_{imp;i,j}- \gamma^s_{imp;i,j})/2\pi T]  \chi_j(m)\tilde{\Delta}_j(m),\\
\chi(n) & = &(2/\sqrt{ \beta_i})\int_{0}^{\infty } dq \exp (-q^2) 
 \nonumber \\
 &   & \times \tan^{-1}(q\sqrt{\beta_i}/(|\tilde{\omega}_i(n)|+i\mu_B 
 H_{c2}\mbox{sgn}(\omega_n)), \\
\beta_i & = & eH_{c2}v_{Fi}^2/2,  \\
\lambda_{i,j}(n) & = & \int_0^{\infty} d\omega \omega\alpha^2_{i,j}F(\omega)/
(\omega^2+\omega_n^2).
\end{eqnarray}
The  bands at $E_F$ are labeled by
 $i$,$j$. Here $\omega_n$=$2\pi T (2n+1)$ are the Matsubara 
frequencies, $\alpha^2_{i}F(\omega)$ and $\tilde{\Delta}_i$, denote the
 spectral density, the superconducting order parameter of
the $i^{th}$ band  respectively. In our approach, as in any two-band model,  
two gaps, below and above the BCS-value of 3.5$k_BT_c$, occur naturally.
 In general,    
interband coupling ($i$$ \neq $$j$) mediated 
by phonons ($\alpha^2_{i,j}F(\omega)$) and impurities 
 is important. Since there is no experimental evidence \cite{schmidt94,suh96} 
for the  presence
of magnetic impurities in high quality  samples,
we neglect the   magnetic  scattering rate $\gamma^s_{imp}$.
For  the quantification of the non-magnetic counterpart 
$\gamma_{imp}\approx 2\pi T_D$, 
the Dingle temperatures, T$_D$,
measured  by the de Haas-van Alphen (dHvA) effect are 
very suitable
\cite{heinecke95,goll96,nguyen96}.
The experimental values $T_D$=2.8 K and 4K reveal  
$\gamma_{imp}$=18K and 25K  for our  YNi$_2$B$_2$C and LuNi$_2$B$_2$C 
single crystals,
 respectively, indicating that the clean limit is 
 reached since $\gamma_{imp} \leq 2\Delta_0 \approx 51 K$ holds for both
 samples, where $2\Delta_0$ denotes the smaller of the two gaps. Hence, the
  scattering by impurities can be neglected
  setting $\gamma_{imp}$=0.  
  In the weak coupling limit of an ISB 
  case, Eqs.\ (1-5) are equivalent to
 the well-known
WHH-theory\cite{werthamer66}. Any  
anisotropy of $H_{c2}$ can be described
by a similar, but  much more tedious, system of equations\cite{langmann92}.
 Since the  measured anisotropy is relatively weak, it will be ignored
  for the sake 
 of simplicity. Therefore, only $H^c_{c2}(T)$'s
  will be compared with those computed for our
   isotropic models.

The standard ISB model \cite{carbotte90}  
describes {\it quantitatively} the renormalization of
the physical properties of metals  due to 
electron-phonon (el-ph) interaction.
The input  parameters of the ISB model are 
the density of states at $E_F$, $N(0)$, the Fermi velocity
$v_F$, the  impurity scattering rate $\gamma_{imp}$, the Coulomb 
pseudopotential
$\mu^{*}$, and the spectral function
 $\alpha^2F(\omega)$ of the  el-ph interaction.
 These quantities can be  determined  
from a few  experimental data: the normal state low-$T$  
electronic specific heat  $\gamma_ST$, 
the plasma frequency $\omega_{pl}$ inferred
from the optical conductivity, $H_{c2}$(0), $T_c$ and its 
isotope exponent $\alpha$, as well as the normal state low-$T$ 
dc resistivity $\rho(0)\approx\rho(T_c)$ which similarly to $T_D$, gives 
a direct measure of 
 the sample purity.   
We adopt  for the Coulomb pseudopotential $\mu^{*}$=0.1 and
 $\hbar\omega_c$=600 meV for the 
energy cutoff in Eq.\ (2).  
 The total el-ph coupling constant,
  $\lambda$=2$\int d\omega \alpha^2F(\omega)/ \omega$,
can be estimated  from the boron isotope effect 
 $\alpha_B\approx$0.2\cite{franck95} and the 
 well-resolved phonon spectrum\cite{renker97}.
 
 We first consider  LuNi$_2$B$_2$C.
To find a lower bound for  
$\lambda$, we  
accounted for only  the high-energy carbon phonons centered at 
50 meV and the boron branch at 100 meV.
Fitting the experimental $\alpha_B$ and T$_c$ values,
we obtained   the partial coupling constants $\lambda_{100}$=0.31, 
$\lambda_{50}$=0.22,  and $\lambda$=$\lambda_{100}+\lambda_{50}$=0.53,
where the subscripts denote the corresponding phonon energies in meV.
 An upper bound of 
$\lambda$=0.77 has been found using the Lu  phonons centered near 
9 meV ($\lambda_{9}$=0.34) and the same B band ($\lambda_{100}$=0.43)
as in the case before. In the following 
a wide averaged spectrum with 
$\lambda$=0.65 ($\lambda_{100}$=0.37,
$\lambda_{50}$=0.12, $\lambda_{9}$=0.16) will be  used  which reproduces
 the experimental values of $\alpha_B$ and $T_c$.
 $N(0)$=11.8 mJ/mol~$k_B^2$K$^2$ has been estimated from
the experimental value \cite{michor95} of   $\gamma_{S}=2\pi^2 k_B^2 (1+\lambda)N(0)/3$=
19.5 mJ/mol K$^2$. The value of $v_F$=2.76$\cdot$10$^7$ cm/s
 follows from the experimental value \cite{bommeli97} of the  plasma frequency 
 $\hbar \omega_{pl}=\sqrt{4\pi e^2 v_F^2 N(0) /3}$ =4.0 eV.
The analogous values for YNi$_2$B$_2$C are
$\lambda$=0.637,
$N(0)$=11.1 mJ/mol $k_B^2$K$^2$, $v_F$=
3$\cdot 10^7$ cm/s and
 $H_{c2}(0)$=2T, where the data of  
  Refs.\ 8 and 34
 have been used.

We solved Eqs.\ (1-5)
 with these parameter sets for two types of 
 spectral densities $\alpha^2F(\omega)$:
(i) a wide averaged spectrum and (ii) 
a single Einstein mode peaked at $\hbar \omega_E$=42.4 meV chosen 
 to yield the experimental $T_{c}$=16.5 K  for LuNi$_2$B$_2$C using
the same value of $\lambda$=0.65  as in the first case.
The results are shown in Fig.\ref{fig1}.
 Note that in the
intermediate coupling regime under consideration, as expected, 
 $H_{c2}(T)$  is 
 insensitive to details of the shape of $\alpha^2F(\omega)$ \cite{carbotte90}
 and, in the clean-limit case, also insensitive to the actual value of the 
  small
  scattering rates.
 Comparing the LuNi$_2$B$_2$C data with the ISB curves one clearly
  realizes
strong deviations. In particular, 
there
is  a discrepancy of about 3 between experimental and ISB model  
values of $H_{c2}$(0). For YNi$_2$B$_2$C the discrepancy reaches even a 
factor of  5.
 Therefore it makes no sense to discuss any
details of the shape of $H_{c2}(T)$, such as the PC, resulting
in deviations of the $H_{c2}(T)$ curves of the order 10 to 20 $\%$ , until
the reason for the large 
failure to account for the magnitude of $H_{c2}(0)$ has been   elucidated.
We remind the reader 
how this serious difficulty was circumvented 
in previous studies.  At first, frequently the quantity
 $h_{c2}(T)=-H_{c2}(T)/$[$T_c($d$H_{c2}/$d$T)_{T=T_c}]$, describing 
 the shape of the $H_{c2}(T)$ curve,
 has been considered, but the {\it absolute} values of $H_{c2}(T)$ 
have not been discussed at all
\cite{metlushko97}.
At second, since  
within the ISB-model $H_{c2}$(0) is a monotonically {\it increasing} function 
of the impurity content, in principle, large $H_{c2}$(0)-values might be 
obtained.
 To check this approach
 \cite{rathnayaka97}
  we calculated
the impurity scattering rate
 $\gamma_{imp}$
which is required to 
increase $H_{c2}$(0) up to the LuNi$_2$B$_2$C value of 7.6 T. Thus we 
obtain
$\gamma_{imp} \approx $ 427 K which would
lead to  $\rho (0) \approx $17
$\mu \Omega$cm which strongly deviates  from the 
experimental value   $\rho (0)$ $\approx$2.5 $\mu \Omega$cm.
In this context we note that our   data and those of Ref.\ 35
show  dependencies just opposite to those predicted by  the ISB-model: 
$H_{c2}(0)$ and $T_{c}$  {\it increase}
 when $\gamma_{imp}$  decreases!
 At third, a  further empirical parameter,  
 the clean limit coherence length, has been introduced in Ref.\ 35. 
  However, this 
 results in the overdetermination of
 the   model parameter set and  
  the consistency of the
 two values
  $v_F$ obtained  using (i) a clean-limit coherence
  length and (ii)  
   normal state data
   has not been checked.
  Thus the ISB approach fails to
 explain simultaneously the three values
 $\gamma_{S}$=19.5 mJ/mol  K$^2$, $\hbar \omega_{pl}$=4 eV, $H_{c2}(0)$=7.6 T
 in the clean limit, and the four values 
 $\gamma_{S}$, $\omega_{pl}$, $H_{c2}(0)$, $\rho$=2.5 $\mu \Omega$cm
 in the dirty limit. In addition, the ISB model is also unable to explain 
  the small gap values
  2$\Delta_0$/$k_BT_{c}<$3.5 observed in microwave \cite{jacobs95},
 tunneling \cite{dewilde97}, and dHvA \cite{terashima97}
  measurements. Furthermore
 the PC of  $H_{c2}(T)$ near $T_c$
 and the extended quasi-linear behavior of $H_{c2}(T)$  
 down to
 $T \sim$ 1 to 2 K cannot
 be described within the ISB.
  The ISB model with  a {\it single} $v_F$
contradicts  the dHvA data which clearly 
 show the presence of at least six different sections $F_{\alpha},...
  F_{\eta}$\cite{goll96,nguyen96} with roughly {\it two} or {\it three}
  groups  of $v_F$'s.
 
  Turning to our TBM 
   we  solved Eqs.\ (1-5). For the sake of 
simplicity, the same phonon spectrum  as in the ISB
case discussed above has been adopted. We achieve
  an excellent   agreement with the LuNi$_2$B$_2$C
 data (see Fig.\ 1)
  for  $\lambda_1$=0.51, $\lambda_2$=$\lambda_{21}$=0.2,
 $\lambda_{12}$=0.4, $\mu^{*}_1$=$\mu^{*}_2$=0.1,
 $v_{F2}$=3.7$\cdot$10$^7$cm/s, $v_{F,1}$=0.96$\cdot$10$^7$cm/s.
     For YNi$_2$B$_2$C we used  the following set: 
    $\lambda_1$=0.5, $\lambda_2$=$\lambda_{21}$=0.2,
 $\lambda_{12}$=0.4,	 $\mu^{*}_1$=$\mu^{*}_2$=0.1,
 $v_{F2}$=3.8$\cdot$10$^7$cm/s, $v_{F1}$=0.85$\cdot$10$^7$cm/s,
  $N(0)$=11 mJ/mol k$_B^2$K$^2$. 	
 Reproducing $T_c$=15.6 K, the adopted values of
 $N(0)$ agrees well  with the 
LDA-value\cite{lee94} of 9.5 mJ/mol k$_B^2$K$^2$. The plasma frequency 
$\hbar\omega_{pl}$
=4.4 eV  
is in accord with  $\hbar\omega_{pl}$=4.25 eV obtained in Ref.\ 34.
From  the obtained $v_F$'s it is concluded that $\omega_{pl}$ is mainly 
related to the second weakly coupled band. Then transport, optical and 
tunneling
data mainly exhibit the properties of that band whereas the strongly coupled 
band remains almost hidden. 
 The calculated value of the penetration depth at
 4.2 K,  100 nm, is in agreement with the data of Ref.\ 35.
 Our $\gamma_S$=17.2 mJ/mol K$^2$
should be compared with   $\gamma_S$=18.2 mJ/mol K$^2$ reported in Ref.\ 8.  
Finally, we arrive at $H_{c2}$(0)=9.4 to 9.9 T
in good agreement  with our experimental 
value $H_{c2}$(0)=10.6 $\pm 0.2$ T.
The experimental  $H_{c2}(T)$ curve of YNi$_2$B$_2$C
 together with results of the TBM
  are shown in 
Fig.\  2.
The values $v_{F} \approx$ 4.2$\cdot$10$^7$cm/s and 0.7 to
1.3$\cdot$10$^7$cm/s 
of the extremal orbits F$_{\beta}$ and F$_{\eta 1/2}$, respectively,
 derived from earlier dHvA
 data
\cite{goll96,nguyen96} on the same YNi$_2$B$_2$C single crystal 
do not  deviate much from the   parameters
 $v_{F2}\approx$3.8 $\cdot$10$^7$cm/s and  $v_{F1} \approx 0.8$ 
 to 0.96$\cdot$10$^7$cm/s introduced
 empirically in our approach.
    The remaining  deviations might be due to
     natural differences between $v_F$'s on extremal orbits
 seen in the
dHvA-experiments and the
 corresponding {\it effective} quantities  
 of our  TBM
  which 
 contains implicitly information on the whole Fermi surface.
 
  Further calculations within our TBM
   reveal that 
the PC of $H_{c2}(T)$ mainly depends  upon the strength of 
 the interband coupling ($\lambda_{12},\lambda_{21}$) and 
 to a lesser extent
 upon the ratio of the Fermi velocities
and the intraband coupling strength ($\lambda_{1},\lambda_{2}$).
These findings may  also be interpreted
in terms of the flat Ni-derived band near E$_F$ and the dispersive 
bands crossing E$_F$ seen in the LDA-band  
 structure \cite{pickett94}.
 The
 strong 
mixing of these bands might 
be viewed as the microscopic origin for
 significant
 interband
  coupling.
 Further  work
is required
 to clarify  this point.
   For  low $T <$4 K, $H_{c2}(T)$ 
       is very sensitive
to the 
\begin{figure}
\psfig{figure=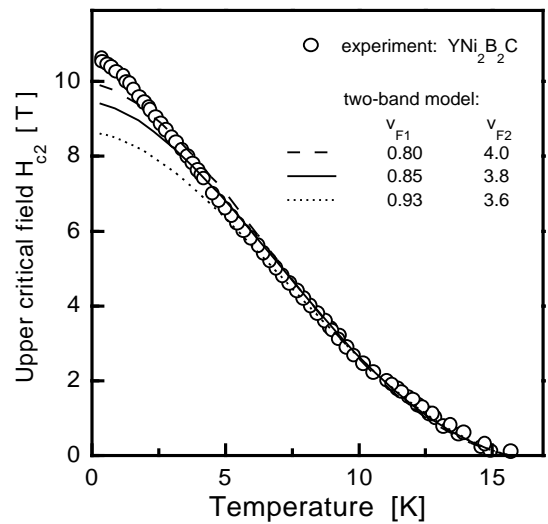,width=8.0cm,height=7.9cm}
\caption{Temperature dependence of $H_{c2}(T)$ for YNi$_2$B$_2$C.
 Experimental points for the magnetic field $\vec{H} \| $ c-axis.
  The Fermi-velocities of the two-band model (TBM)
   are given in units of
 10$^{7}$ cm/s. For the interaction constants see the text. 
}
 \label{fig2}
 \end{figure}
 \noindent 
 actual   $v_{F,2}/v_{F,1}$ ratio. 
This is illustrated by the 
additional curves  shown in Fig.\ 2.  The variation of 
  $v_{F1}$ and $v_{F2}$ results in
  slightly higher  and lower $H_{c2}$(0)
  values, respectively. 
    Our empirical
 parameter sets  for  LuNi$_2$B$_2$C 
   and YNi$_2$B$_2$C
  differ  almost
    only in their $v_{F1}$ values which
     roughly scale
with   the 
Ni-Ni distance  as $d_{Ni-Ni}^{-5}$.
In this   context the study of  Sc(Th)Ni$_2$B$_2$C crystals
having much reduced (increased) Ni-Ni distances is
 of interest.

Our analysis reveals that both compounds can be well
 described within an effective two-band
 model  provided there are at least 
 two groups of electrons having (i) 
significantly
 different Fermi-velocities, (ii) strong coupling in the 
small-$v_{F}$
band, as well as (iii) sizeable coupling between the small-$v_{F}$ 
and the large-$v_{F}$
 band.
This case  differs from   the situation considered in Refs.\ 24 and 26.  
  There, the stronger coupling
 is in the large-$v_{F}$ band and the
curvature of $H_{c2}(T)$ near $T_c$ is {\it negative}. A
PC would only appear  at intermediate 
$T$  if the interband coupling and the impurity scattering are 
both weak. In contrast, in this region $H_{c2}(T)$ shows  almost no curvature
 in our model.
In other words, the result of Refs.\ 24 and 26
can be understood as an average over 
 two weakly coupled superconductors, the first with a 
 high $H_{c2}(0)$ but a low $T_{c}$  and the second one with
a small $H_{c2}(0)$ but high $T_{c}$. In our case
the isolated 
small-$v_{F}$  subsystem would have   high values of $\lambda_, H_{c2}(0)$,
 and $T_{c}$. The values of $ H_{c2}(0)$ and $T_{c}$ of the coupled 
 system are reduced by the second large-$v_{F}$ subsystem with weak
 interaction parameters. In this case, which to the best of our knowledge has
  not been  considered so far,
 the PC of the resulting $H_{c2}(T)$ near $T_c$ becomes
 a direct manifestation of that interband coupling. In our TBM, 
  the PC of 
 $ H_{c2}(T)$, as well as $ H_{c2}(0)$, 
  and  $T_c$ are
   suppressed by growing
impurity content and the PC  vanishes  upon 
reaching the dirty limit with $T_c \approx$ 11 K. 
 Thus the observed maximal PC of $H_{c2}(T)$ near $T_c$
of multi-band systems of the type under consideration 
 can be regarded as 
a direct measure of the high sample quality, as opposed to the widely
 spread 
belief attributing it simply to 
 sample inhomogeneities.
  The latter scenario
  is excluded by the sharp transition 
  in the $\chi(T)$ data.

To summarize, we have shown that 
 the superconductivity
 of  pure non-magnetic borocarbides
 should be described within a multi-band picture. The
 isotropic TBM
  gives a reasonable starting point
 toward the understanding of the  mechanism of superconductivity in these
 compounds.
 Combined studies of quantum oscillations (dHvA) and 
 $H_{c2}(T)$
unified by Eliashberg analysis are found to be valuable
supplementary tools to elucidate the specific role of  
subgroups of electrons having small 
$v_F$'s and  
strong coupling  to bosons. These electrons may be readily
 overlooked 
 by 
 other experimental techniques. 

We thank 
  O.\ Dolgov, D.\ Rainer, 
  E.\ Maksimov, 
  H.\ Braun, 
  N.\ Schopohl,  
 M.\ Golden, 
J.\ Fink, L.\ Schultz, 
H.\ Eschrig,  
 and A.\ Gladun 
for discussions. 
This work was  supported by the INTAS-93-2154 grant, 
the SFB 463,  and the Deutsche Forschungsgemeinschaft.


\begin{references}
\bibitem[*]{shu}
 On leave from 
 Inst.\ of Spectroscopy RAS, 142092 Troitsk, Russia.
 E-mail: shulga@fly.triniti.troitsk.ru
 \bibitem[+]{dre}
 Corresp.\ author, E-mail: drechsler@ifw-dresden.de
  \bibitem{cava94} R.J.\ Cava {\it et al.,} Nature (London) 
{\bf 367}, 146, 252 (1994).
\bibitem{nagarajan94} R.\ Nagarajan {\it et al.,} Phys.\ Rev.\ Lett.\ {\bf 72}, 
274 (1994).
\bibitem{cho95} B.K.\ Cho {\it et al.,} Phys.\ Rev.\ B 
{\bf 52}, R3844 (1995).
\bibitem{evertsmann96} K.\ Eversmann, {\it et al.,} Physica  C
 {\bf 266}, 27 (1996).
 \bibitem{fuchs95} K.-H.\ M\"uller {\it et al.,}
  J.\ Appl.\ Phys.\ {\bf 81}, 4240 (1997).
\bibitem{lai95} C.C\ Lai, {\it et al.,} Phys.\ Rev.\ B {\bf 51}, 420 (1995).
\bibitem{metlushko97} V.\ Metlushko {\it et al.,} Phys.\ Rev.\ Lett.\ {\bf 79}, 
1738 (1997).
\bibitem{michor95} H.\ Michor {\it et al.,} Phys.\ Rev.\ B {\bf 52},
 16165 (1995).
 \bibitem{rathnayaka97} K.D.D.\ Rathnayaka  {\it et al.,}
 Phys.\ Rev.\ B {\bf 55}, 8506 (1997).
\bibitem{dewilde97} Y.\ De Wilde {\it et al.,}  Phys.\ Rev.\ Lett.\ 
{\bf 78}, 4273 (1997).
 \bibitem{bahcall95} S.R.\ Bahcall,  Phys.\ Rev.\ Lett.\ {\bf 75}, 1376 (1995).
\bibitem{kotliar96} G.\ Kotliar {\it et al.,} Phys.\ Rev.\ Lett.\ {\bf 77}, 
2296 (1996).
\bibitem{alexandrov93} A.S.\ Alexandrov, Phys.\ Rev.\ B {\bf 48}, 10571 (1993).
 \bibitem{pickett94} W.E.\ Pickett {\it et al.,} Phys.\ Rev.\ Lett.\ 
{\bf 72}, 3702 (1994).
\bibitem{lee94} J.I.\ Lee {\it et al.,}  Phys.\ Rev.\ B {\bf 50},
 4030 (1994).
 \bibitem{abrikosov97} A.A\ Abrikosov, Phys.\ Rev.\ B {\bf 56}, 
5112 (1997).
 \bibitem{heinecke95} M.\ Heinecke and K.\ Winzer, Z.\ Phys.\ B {\bf 98}, 147 (1995).
 \bibitem{goll96} G.\ Goll {\it et al.}, Phys.\ Rev.\ B {\bf 53}, R8871 (1996).
\bibitem{nguyen96} L.H.\ Nguyen {\it et al.}, J.\  Low Temp.\ Phys.\ 
 {\bf 105}, 1653 (1996).
 \bibitem{remark0}Using the most complete dHvA data  available for 
 YNi$_2$B$_2$C, only,
 the mean free path  $l_q$=$\hbar v_F
 /2\pi k_B T_D$ 
 for electrons belonging
 to different cross sections 
 can be estimated as 181, 86, 56, and 30 nm, where the  Fermi velocities 
 $v_F$ of
 4.2, 2, 1.3, and 0.7$\cdot 10^7$cm/s, respectively, and the Dingle temperature 
 $T_D$=2.8 K have been used. From  
 $\rho$(0)$\approx \rho_n$=2.5$\mu \Omega$cm, $\l_{\rho}=4\pi v_F/
 \omega_{pl}^2\rho(0)$ only the mean free paths of the large $v_F$ bands
  can be 
 estimated as
  $l_{\rho}\approx $41 (45), where  
$\omega_{pl}$=4.25 (4.0) eV and $v_F$=3.6 (3.7)$\cdot 10^7$ cm/s have been
 used.
All  estimated $l$'s exceed significantly
the GL-coherence 
 length 
 $\xi_0=\sqrt{\Phi_0/2\pi H_{c2}(0)}$=5.5 (6.5) nm for
  YNi$_2$B$_2$C (LuNi$_2$B$_2$C). 
 \bibitem{remark2}
Note 
 that the term ``two-band''  should not be
  taken too literally. Within the Eliashberg
   theory\cite{prohammer87,entel76,allen76,langmann92} a single anisotropic
   band model is equivalent 
   to an isotropic multi-band model  more suitable for theoretical studies.
 \bibitem{moskalenko73}H.\ Suhl {\it et al.}, Phys.\ Rev.\ Lett.\ {\bf 3}, 
552, (1959).
\bibitem{prohammer87}  M.\ Prohammer {\it et al.}, Phys.\ Rev.\ B {\bf 36},
 8353 (1987).
\bibitem{entel76} P.\ Entel {\it et al.}, 
J.\ of Low Temp.\ Phys.\  {\bf 22},
 613 (1976).
\bibitem{allen76} P.B.\ Allen, Phys.\ Rev.\ B {\bf 13}, 1416 (1976).
\bibitem{langmann92} E.\ Langmann, Phys.\ Rev.\ B {\bf 46}, 9104 (1992).
\bibitem{carbotte90} J.P.\ Carbotte, Rev.\ Mod.\ Phys.\ {\bf 62}, 1027
 (1990).
\bibitem{schmidt94} H.\ Schmidt {\it et al.,} Physica C {\bf 235-240},
 779 (1994).
\bibitem{suh96}  B.J.\ Suh {\it et al.,} Phys.\ Rev.\ B. \ {\bf 53} 
R6022 (1996).
\bibitem{werthamer66} N.R.\ Werthamer {\it et al.,}
 Phys.\ Rev.\  {\bf 147}, 295 (1966).
\bibitem{franck95} D.D.\ Lawrie {\it et al.,}  Physica C {\bf 245}, 159 (1995).
\bibitem{renker97} F.\ Gompf {\it et al.,} Phys.\ Rev.\ B {\bf 55},9058
 (1997).
\bibitem{bommeli97} F.\ Bommeli {\it et al.,} Phys.\ Rev.\ Lett.\ {\bf 78}, 
547 (1997).
\bibitem{widder95} K.\ Widder {\it et al.}, Europhys.\ Lett.\ {\bf 30},
 55 (1995).
\bibitem{schmidt97} H.\ Schmidt {\it et al.,}
 Phys.\ Rev.\ B {\bf 55}, 8497 (1997).
 \bibitem{jacobs95} T.\ Jacobs  {\it et al.,}
 Phys.\ Rev.\ B {\bf 52}, R7022 (1995).
\bibitem{terashima97} T.\ Terashima {\it et al.,} Phys.\ Rev.\ B {\bf 56}, 5120
 (1997).
\end{references}
\end{document}